\documentclass{kluwer}
\usepackage{epsfig}
\begin{document}
\begin{article}
\begin{opening}
\title{New values of gravitational moments $J_{2}$ and $J_{4}$ deduced from helioseismology}

\author{R. \surname{Mecheri}}
\author{T. \surname{Abdelatif}}
\author{A. \surname{Irbah}}

\institute{C.R.A.A.G - \emph{Observatoire d'Alger} BP 63 Bouzareah, Alger, Algerie.}

\author{J. \surname{Provost}}
\author{G. \surname{Berthomieu}}

\institute{Departement Cassini, UMR CNRS 6529 - \emph{Observatoire de la C\^{o}te d'Azur, BP 4229, 06304 Nice CEDEX4,
France.}}

\runningtitle{New values of gravitational moments $J_{2}$ and $J_{4}$ deduced from helioseismology}

\runningauthor{Mecheri et al.}

\begin{abstract}
By applying the theory of slowly rotating stars to the Sun, the solar quadrupole and octopole moments $J_{2}$ and
$J_{4}$ were computed using a solar model obtained from \emph{CESAM} stellar evolution code (Morel (1997)) combined
with a recent model of solar differential rotation deduced from helioseismology (Corbard \emph{et al.} (2002)). This
model takes into account a near-surface radial gradient of rotation which was inferred and quantified from \emph{MDI
f-modes} observations by Corbard and Thompson (2002). The effect of this observational near-surface gradient on the
theoretical values of the surface parameters $J_{2}$, $J_{4}$ is investigated. The results show that the octopole
moment $J_{4}$ is much more sensitive than the quadrupole moment $J_{2}$ to the subsurface radial gradient of rotation.
\end{abstract}

\end{opening}

\section{Introduction}
Several theoretical determinations of the $J_{2}$ and $J_{4}$ gravitational moments have been undertaken in case of
different solar differential rotation laws : (i) only radius dependence (Goldreich and Schubert, 1968; Patern\`{o},
Sofia, and Di Mauro, 1996), (ii) quadratic expansion in colatitude cosine terms (Ulrich and Hawkins (1981a and 1981b)),
(iii) angular velocity distribution with a slowly latitude variation determined by mean of helioseimology technics
(Gough (1982)). More recent determinations are those performed by : (i) Armstrong and Kuhn (1999) using a quadratic
expansion rotation law with coefficients obtained by fitting higher resolution helioseismic interior rotation data from
MDI (Scherrer \emph{et al.} (1995)), (ii) Godier and Rozelot (1999) and (iii) Roxburgh (2001) using the differential
rotation model given by Kosovichev (1996) from \emph{BBSO p-modes} observations. This model takes into account the
presence of a constant near-surface radial gradient based on the assumption that the angular momentum is preserved in
the supergranulation layer. The aim of the present work is a contribution to $J_{2}$ and $J_{4}$ determinations using a
new analytical model of solar differential rotation provided by Corbard \emph{et al.} (2002) which has a latitudinal
dependent profile of the
near-surface radial gradient of rotation.\\
If we consider the Sun as an axial symmetry distribution of matter in rotation, the outer gravitational field
$\phi_{out}$ is expressed as :
\begin{equation}\
  \phi_{out}(r,\theta)=-\frac{GM_{\odot}}{r}\left[1-\sum_{n=1}^{\infty}\left(\frac{R_{\odot}}
  {r}\right)^{2n}J_{2n}P_{2n}(\cos\theta)\right]
\end{equation}
where $J_{2n}$ are the gravitational moments, $P_{2n}$ the Legendre polynomials and $r$, $\theta$ respectively the
distance from the Sun centre and the angle to the symmetry axis (colatitude).\\
Since the solar rotation is slow, it induces small perturbations around the spherical equilibrium. These perturbations
can be expanded on Legendre polynomial basis. The distribution of the gravitational potential in the Sun can be written
:
\begin{eqnarray}\
  \phi(r,\theta)=\phi_{0}(r)+\phi_{1}(r,\theta)=\phi_{0}(r)+\sum_{n=1}^{\infty}\phi_{12n}(r)P_{2n}(u)
\end{eqnarray}
where $u=cos\theta$.\\
The gravitational moments $J_{2n}$ are obtained assuming the continuity of the gravitational potential at the surface :
\begin{equation}\
  J_{2n}=\frac{R_{\odot}}{GM_{\odot}}\phi_{12n}(R_{\odot})
\end{equation}
The perturbed potential is obtained by linearization of the equation of hydrostatic equilibrium and the Poisson
equation, leading to :
\begin{eqnarray}
  \frac{d^{2}\phi_{12n}}{dr^{2}}&+&\frac{2}{r}\frac{d\phi_{12n}}{dr}-\left(2n(2n+1)+UV\right)
  \frac{\phi_{12n}}{r^{2}}=U[(V+2)B_{2n}+\nonumber\\&+&
  r\frac{dB_{2n}}{dr}+\frac{4n+1}{2}\int_{-1}^{+1}(1-u^{2})P_{2n}(u)\Omega(r,u)^{2}du]
\end{eqnarray}
where $U=4\pi\rho_{0}r^{3}/M_{r}$, $V=dln\rho_{0}/dlnr$, $M_{r}$
is the mass contained in a sphere of radius $r$ and $\Omega(r,u)$
the angular velocity. $B_{2n}$ is given by:
\begin{equation}\
  ~~~~~~~~B_{2n}(r)=-\frac{1}{2n!}\frac{4n+1}{2^{2n+1}}\int_{-1}^{+1}u
  \Omega(r,u)^{2}\frac{d^{2n-1}}{du^{2n-1}}((u^{2}-1)^{2n})du
\end{equation}
Equation (3) is integrated with the usual boundary conditions,
using $U$ and $V$ provided by a solar model obtained from the
\emph{CESAM} stellar evolution code (Morel (1997)) and a rotation
law derived from helioseismology.

\section{Analytical model of solar differential rotation}

We consider the recent Corbard \emph{et al.} (2002) model of solar differential rotation and, for comparison, the
Kosovichev one (1996) already used by Roxburgh (2001) and Godier and Rozelot (1999). The Corbard model contains a
near-surface radial gradient of rotation inferred from the radial dependence of the \emph{MDI f-modes} observations
(Corbard and Thompson (2002)). Two estimations of this gradient have been derived from different sets of modes leading
to two rotation models denoted afterward by (a) and (b). As for Kosovichev's model, the surface rotation is forced to
surface plasma observations (Snodgrass, 1992).\\
Figure 1 shows the solar rotation profiles corresponding to these models computed for different latitudes. Kosovichev's
model presents a negative constant subsurface radial gradient. The Corbard models have a negative value of the radial
gradient at low latitude which is twice smaller than Kosovichev ones. At high latitude, the gradients are positive with
larger magnitudes for the Corbard model (b).\\
\begin{figure}[b]
\centerline{\epsfig{file=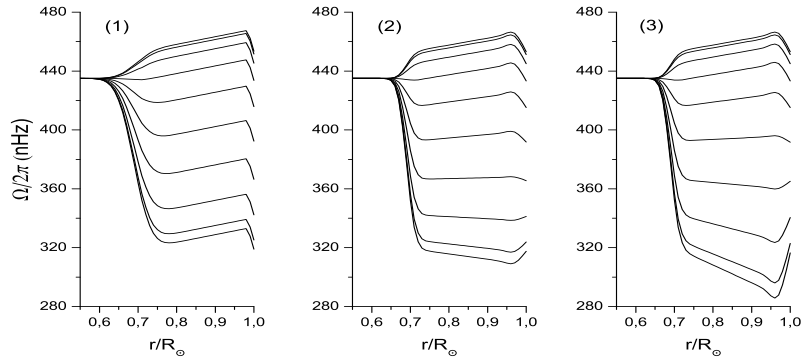,width=4.6in,height=2.18in}}
\caption{Profiles of the solar rotation from $0.55R_{\odot}$ to
the surface for different latitudes computed each $10^{\circ}$
from the Equator (top) to the Pole (bottom).\textbf{(1) Model of
Kosovichev}.~(\textbf{2) Model of Corbard (a)}.~\textbf{(3) Model
of Corbard (b)}.}
\end{figure}
We use an analytical expression derived respectively by Corbard
\emph{et al.} (2002) and Dikpati \emph{et al.} (2002) for the
Corbard rotation laws and for the Kosovichev one. We recall
hereafter the full set of equations and parameters they give for
these rotation laws :
\begin{eqnarray}\
  \Omega(r,u)=A_{1}(r,u)+\Psi_{tac}(r)\left(\Omega_{cz}-\Omega_{0}+a_{2}u^{2}+a_{4}u^{4}\right)
\end{eqnarray}
where
\begin{eqnarray}\
  A_{1}(r,u) & = & \Omega_{0}+\Psi_{cz}(r)\left\{\alpha(u)(r-r_{cz})\right\}+\nonumber\\\
             &+&\Psi_{s}(r)\left\{\Omega_{eq}-\Omega_{cz}-\beta(u)(r-R_{\odot})-\alpha(u)(r-r_{cz})\right\}\
\end{eqnarray}
with
\begin{eqnarray}\
   \alpha(u) & = & \frac{\Omega_{eq}-\Omega_{cz}+\beta(u)(R_{\odot}-r_{s})}{r_{s}-r_{cz}}\nonumber
\end{eqnarray}
$\Omega_{0}$, $\Omega_{eq}$ and $\Omega_{cz}$ are respectively the
constant rotation of the radiative interior zone, the equatorial
rate at the surface and at the top $r_{cz}$ of the tachocline
localized at $r_{tac}$. The $a_{2}$ and $a_{4}$ constants describe
the latitudinal differential rotation. $\beta(u)$ represents the
latitudinal dependance of the rotation radial gradient below the
surface down to a radius $r_{s}$. This gradient depends on the
latitude through the $\beta_{0}$, $\beta_{3}$ and $\beta_{6}$
constants by the Equation :
$\beta(u)=\beta_{0}+\beta_{3}u^{3}+\beta_{6}u^{6}$.

The $\Psi_{_{x}}$ function where $x$ stands for $tac$, $cz$ or
$s$, models the transition between different gradients. An error
function centered at $r_{x}$ with width $\omega_{x}$ is used for
this goal :
$\Psi_{x}(r)=0.5\left(1+erf\left[2(r-r_{x})/\omega_{x}\right]\right)$.\\
All the rotation laws have the following common parameters :
$\Omega_{0}=435nHz$, $\Omega_{eq}=452.5nHz$,
$\Omega_{cz}=453.5nHz$, $r_{tac}= 0.69R_{\odot}$, $r_{cz}=
0.71R_{\odot}$, $a_{2}=-61nHz$, $a_{4}=-73.5nHz$. The parameters which are different for the three laws are given in Table I.\\

\begin{table*}[h]
\begin{tabular}{c c c c c c c c}\
  model & $\omega_{tac}/R_{\odot}$ & $\omega_{cz}/R_{\odot}$ & $\omega_{s}/R_{\odot}$ & $r_{s}/R_{\odot}$
  & $\beta_{0}$ & $\beta_{3}$ & $\beta_{6}$\\\hline
  Kosovichev (1996) & 0.1 & 0 & 0 & 0.983 & 891.5 & 0 & 0\\
  Corbard (a) (2002) & 0.05 & 0.05 & 0.05 & 0.97 & 437 & -214 & -503\\
  Corbard (b) (2002) & 0.05 & 0.05 & 0.05 & 0.97 & 437 & 0 & -1445\\
\end{tabular}
\caption[]{The non common parameter values between the rotation models. The  $\beta_{0}$, $\beta_{3}$ and $\beta_{6}$
parameters are given in $nHz/R_{\odot}$.} \label{parset}
\end{table*}

\section{Results and discussion}

We present in Table II the computed values of $J_{2}$ and $J_{4}$ obtained with different solar rotation models
described in Section 2 and with an uniform rotation equal to the rotation rate of the solar radiative zone
$\Omega_{0}$, for comparison. Table III gives also some values of $J_{2}$ and $J_{4}$ presented by other authors.\\
Our results show that the differential rotation in the convective zone reduces $J_{2}$ value of about $0.8\%$ when the
Corbard models are considered and about $0.5\%$ in the case of Kosovichev's model. For this last case, our $J_{2}$
determination is larger than the value found by Godier and Rozelot (1999) but in agreement with the one obtained by
Roxburgh (2001), both using Kosovichev's model. Our values are also in agreement with those given by Patern\`{o}
\emph{et al.} (1996), Pijpers (1998) and Armstrong and Kuhn (1999) (see Table III). All these values deviate from the
range obtained by Ulrich and Hawkins (1981a and 1981b) in the case of the rotation law defined as a simple quadratic
expansion. The difference between the subsurface radial gradients induces only a small reduction on $J_{2}$ values. It
is about $0.25\%$ between Kosovichev's model and Corbard's ones. This difference is however about $0.1\%$ between the
two Corbard models.\\
\begin{table*}
\begin{tabular}{c c c c c}
  \hline Model of rotation & $J_{2}(\times10^{-7})$ &
  $J_{4}(\times10^{-9})$ \\\hline
  Uniform rotation ($\Omega_{0}$) & 2.217 & 0 \\ 
  Kosovichev (1996) & 2.205 & -4.455 \\ 
  Corbard (a) (2002) & 2.201 & -5.601 \\ 
  Corbard (b) (2002) & 2.198 & -4.805 \\ 
\end{tabular}
\caption[]{The $J_{2}$ and $J_{4}$ values corresponding to the different rotation models. $\Omega_{0}=2.733~\mu rd/s$
is the rotation in the radiative zone.}\label{parset}
\end{table*}
\begin{table*}[b]
\begin{tabular}{c c c}
  \hline Authors & $J_{2}(\times10^{-7})$ &
  $J_{4}(\times10^{-9})$ \\\hline
  Ulrich and Hawkins (1981) & $1.0 < J_{2}< 1.5$ & $2.0<|J_{4}|<5.0$ \\ 
  Gough (1982) & $36$ & - \\ 
  Patern\`{o} \emph{et al.} (1996) & $2.22$ & - \\ 
  Pijpers (1998) & $2.18$ & - \\ 
  Godier and Rozelot (1999) & $1.6$ & - \\ 
  Armstrong and Kuhn (1999) & $2.22$ & $-3.84$ \\ 
  Roxburgh (2001) & $2.206$ & $-4.45$ \\ 
\end{tabular}
\caption[]{Some computed values of $J_{2}$ and $J_{4}$ obtained by other authors. The large value of Gough (1982) is
due to an estimation of the internal rotation deduced from earlier helioseismic observations}\label{parset}
\end{table*}
As expected, the effect of the subsurface radial gradient is more
important on $J_{4}$ gravitational moment. $J_{4}$ absolute values
obtained using the models of Corbard (a) and (b) are respectively
about 20\% and 7\% larger than the one obtained with Kosovichev's
model. The (a) Corbard model increases the $J_{4}$ absolute value
of about 14\% compared to the one obtained from the (b) Corbard
model. Our $|J_{4}|$ value corresponding to Kosovichev's model is
in agreement with the one given by Roxburgh (2001) using the same
model. However, those obtained from Corbard's models are larger
than other values given in Table III. All these values are
consistent with the range given by Ulrich and Hawkins (1981a and
1981b) except for the (a) Corbard model.\\
Rotation induces a distortion of the solar surface which can be
roughly related to $J_{2}$ through the following quantity often
called oblateness :
\begin{equation}
  \frac{R_{e}-R_{p}}{R_{\odot}}\approx\frac{3}{2}J_{2}+\frac{\Omega^{2}_{s}R_{\odot}^{3}}{2GM_{\odot}}
\end{equation}
where $\Omega_{s}$ is an effective rotation rate. $R_{e}$, $R_{p}$
and $R_{\odot}$ are respectively the equatorial, polar and mean
solar radius. This formula is strictly valid for an uniform
rotation or for a rotation constant on cylinders. For a solar
rotation which presents a complex profile not constant on
cylinders, Patern\`{o}, Sofia, and Di Mauro (1996) proposed an
expression of $\Omega_{s}$ derived from the surface rotation
$\Omega(R_{\odot},u)$. In our case, $\Omega_{s}$ will be the same
for the three models since they are built so as to have the same
surface rotation. Thus, in this rough description, the effect of
different subsurface radial gradients of rotation on the
oblateness appears through the modification of the $J_{2}$
gravitational moment. It is negligible since the main term in
Equation (8) is the surface rotation term. The value of the
oblateness found is $9.1\times10^{-6}$. It is in agreement with
observations of Lydon and Sofia (1996) and Rozelot and R\"{o}sch
(1997) but slightly larger than those of Kuhn \emph{et al.} (1998)
and Armstrong and Kuhn (1999). Rozelot, Godier, and Lefebvre
(2001) presented new developments taking into account the surface
latitudinal differential rotation to link $J_{2}$ and $J_{4}$ to
the solar equatorial and polar radius. To the lowest order, their
formula reduces to Equation (8) with another definition of
$\Omega_{s}$. For our surface rotation, their formula leads to an
oblateness value equal to $11.4\times10^{-6}$. We have estimated
that the effects of the second terms are of the order of (4/1000)
in the case of Kosovichev's model. New constraints on the
oblateness and the shape of the solar surface will hopefully be
provided by the future \emph{PICARD} microsatellite
\emph{CNES-mission} (Thuillier \emph{et al.} (2003)).\\
In conclusion, the octopole moment $J_{4}$ is much more sensitive than the quadrupole moment $J_{2}$ to the inclusion
of the latitudinal and radial differential rotation in the convective zone and particularly to the subsurface radial
gradient of rotation. Indeed, an important subsurface radial gradient at high latitude decreases significantly the
value of $|J_{4}|$ while it does not affect significantly the $J_{2}$ value.

\begin{acknowledgements}

We thank Thierry Corbard for providing his model of rotation and the referee for his constructive remarks and comments.
This work has been performed with support of the French Foreign Affaire Ministry in the framework of scientific
cooperation between France and Algeria (contract 00MDU501)

\end{acknowledgements}

\end{article}
\end{document}